\documentclass[12pt]{article}
\def\1{\'{\i}}

\usepackage[cm]{fullpage}
\usepackage{amssymb}

\title{Approximate Kerr-like Metric with Quadrupole}
\author{Francisco Frutos Alfaro \\
{\small School of Physics and Space Research Center 
of the University of Costa Rica}}
\date{\today}

\begin{document}
\maketitle

\abstract{
A new approximate metric representing the spacetime of a rotating deformed 
body is obtained by perturbing the Kerr metric to include up to the second 
order of the quadrupole moment. It has a simple form, because it is Kerr-like. 
Its Taylor expansion form coincides with second order quadrupole metrics 
with slow rotation already found. Moreover, it can be transformed to an 
improved Hartle-Thorne metric, which guarantees its validity to be useful in 
studying compact object, and that it is possible to find an inner solution.}

\section{Introduction}
\label{sec:Intro}

\noindent
Nowadays, it is widely believed that the Kerr metric does not represent the 
spacetime of a rotating astrophysical object. It seems that there is no 
reasonable perfect fluid inner solution which serves as source of this 
spacetime \cite{Hernandez}. Moreover, the relationship between its multipole 
moments and its angular momentum may not represent correctly the external 
field of any realistic stars \cite{Thorne}.

\noindent
The Ernst formalism \cite{Ernst} and the Hoenselaers-Kinnersley-Xanthopoulos 
(HKX) transformations \cite{HKX} are very useful to find exact axial solutions 
of the Einstein field equations (EFE). These formalisms allow to include 
desirable characteristics (rotation, multipole moments, magnetic dipole, etc.) 
to a given seed metrics. In this article, we develop a perturbative method by 
means of the Lewis metric \cite{Carmeli} to find solutions with quadrupole 
moment, using the Kerr spacetime as seed metric. Our method consists in 
modifying four potential functions of the Lewis metric and maintaining 
the cross term potential function. This method was applied successfully in 
obtaining other approximative metrics \cite{Frutos1,Frutos2,Montero}. 

\noindent
To ensure the validity of a metric, the given metric is expanded to its 
post-linear form and compared with the post-linear version of the 
Hartle-Thorne (HT) spacetime \cite{HT,Quevedo}. The reason is that it is 
possible to find an inner solution corresponding to the HT metric 
\cite{Boshkayev}. This new approximation can be considered as an improvement of 
the HT spacetime, because it has spin octupole and the HT has not this one. 
There are several exact metrics \cite{Manko,QM}, but these ones are more 
appropriate for numerical works. This Kerr-like metric has a simple form and 
can be useful for theoretical works. For instance, it may used to investigate 
the influence of the mass quadrupole in the light propagation and the light 
cone structure of this Kerr-like spacetime \cite{FrutGrave}. 

\noindent
This paper is organized as follows. Our perturbation method of the Kerr metric 
using the Lewis one is discussed in section 2. In section 3, it is shown that 
the application of this method leads to a new approximate solution to the EFE 
with rotation and quadrupole moment. It is checked by means of a REDUCE 
program that the resulting metric is a solution of the EFE \cite{Hearn}, and
this program is available upon request. In section 4, the exterior HT metric is 
briefly explained and compared to our Kerr-like ones. We also compare it with 
the Erez-Rosen (ER) metric \cite{Carmeli} without rotation. The comparison of 
our metric to the HT spacetime assures that our metric has astrophysical 
meaning. A comparison with other stationary metrics is given in section 5. 
A summary and discussion of the results is presented in section 6.


\section{The Perturbing Method for the Kerr Metric}
\label{sec:PKM}

\noindent 
First of all, we need a spacetime to work on. To this end, the Lewis metric is 
chosen and is given by \cite{Carmeli} 

\begin{equation}
\label{lewis} 
{d}{s}^2 = - V d t^2 + 2 W d t d \phi + X d \rho^2 + Y d z^2 + Z d \phi^2 ,
\end{equation}

\noindent 
where the chosen canonical co\-or\-di\-na\-tes are $ x^{1} = \rho $ and 
$ x^{2} = z $. The potentials $ V, \, W, \, Z $, $ X = {\rm e}^{\mu} $ and 
$ Y = {\rm e}^{\nu} $ are functions of $ \rho $ and $ z $ with 
$ \rho^2 = V Z + W^2 $. Taking $ \mu = \nu $, performing the following changes 
of potentials $ V = f , \; W = \omega f , \; Z = {\rho^2}/{f} - \omega^2 f $ 
and choosing $ {\rm e}^{\mu} = {{\rm e}^{\gamma}}/{f} $, one get the 
Weyl-Papapetrou metric \cite{Carmeli}

\begin{equation}
\label{papapetrou} 
{d}{s}^2 = - f (d t - \omega d \phi)^2 
+ \frac{{\rm e}^{\gamma}}{f} [d \rho^2 + d z^2] + \frac{\rho^2}{f} d \phi^2 .
\end{equation}

\noindent 
The Ernst formalism and HKX transformation are based on this metric. 
Here, these formalisms are not employ to generate a new one. Rather, 
a new method to find a Kerr-like metric with quadrupole is developed. 
To this goal, we use the known transformation that leads to the Kerr metric 
\cite{Carmeli}

\begin{equation}
\label{chandra} 
\rho = \sqrt{\Delta} \sin{\theta} \qquad {\rm and} 
\qquad z = (r - M) \cos{\theta} ,
\end{equation}

\noindent 
where $ \Delta = r^2 - 2 M r + a^2 $ 
(with $ M $ as the mass of the object and $ a $ as the rotation parameter).

\noindent 
Now, one chooses the Lewis potentials as follows 

\begin{eqnarray}
\label{potential}
V & = & V_K \, {\rm e}^{- 2 \psi} 
= \frac{1}{{\tilde{\rho}}^2} [\Delta - a^2 \sin^2{\theta}] \, {\rm e}^{- 2 \psi} 
\nonumber \\
W & = & - \frac{2 J r}{{\tilde{\rho}}^2} \sin^2{\theta} \nonumber \\
X & = & X_K {\rm e}^{2 \chi} = {\tilde{\rho}}^2 \frac{{\rm e}^{2 \chi}}{\Delta} \\
Y & = & Y_K {\rm e}^{2 \chi} = {\tilde{\rho}}^2 \, {\rm e}^{2 \chi} \nonumber \\
Z & = & Z_K {\rm e}^{2 \psi} = \frac{\sin^2{\theta}}{{\tilde{\rho}}^2} 
[(r^2 + a^2)^2 - a^2 \Delta \sin^2{\theta}] \, {\rm e}^{2 \psi} , \nonumber 
\end{eqnarray}

\noindent 
where $ J = M a $. The potentials $ V_K, \, W, \, X_K, \, Y_K, \, Z_K $ are 
the Lewis potentials for the Kerr spacetime, and 
$ {\tilde{\rho}}^2 = r^2 + a^2 \cos^2{\theta} . $

\noindent 
The cross term potential $ W $ is unaltered to preserve the following metric 
form

\begin{eqnarray} 
\label{superkerr}
d{s}^2 & = & - \frac{\Delta}{{\tilde{\rho}}^2} 
[{\rm e}^{- \psi} dt - a {\rm e}^{\psi} \sin^2{\theta} d \phi]^2
+ \frac{\sin^2{\theta}}{{\tilde{\rho}}^2} 
[(r^2 + a^2) {\rm e}^{\psi} d \phi - a {\rm e}^{- \psi} d t ]^2 \\
& + & {\tilde{\rho}}^2 {\rm e}^{2 \chi} \left(\frac{d r^2}{\Delta} 
+ d \theta^2 \right) .
\nonumber 
\end{eqnarray}

\noindent 
The so chosen potentials guarantee that one gets 
the Kerr metric if $ \psi = \chi = 0 $. The function $ \psi $ and $ \chi $ 
will be determine approximatively from the EFE. 

\newpage

\section{The Approximative Kerr Metric with Quadrupole}
\label{sec:super}

\noindent 
Now, we have to solve the EFE perturbatively

\begin{equation}
\label{einstein} 
G_{i j} = R_{i j} - \frac{R}{2} g_{i j} = 0 ,
\end{equation}

\noindent 
where $ G_{i j} $ ($ i, \, j = 0, \, 1,\, 2, \, 3 $) are the Einstein tensor 
components, $ R_{i j} $ are the Ricci tensor components, and $ R $ is the 
curvature scalar. 

\noindent
Terms such as

\begin{eqnarray}
\label{condition}
M^2 \frac{\partial \sigma}{\partial x^i} & \sim 0 , \qquad 
M \displaystyle{\left(\frac{\partial \sigma}{\partial x^i} \right)^2} & 
\sim 0 , \nonumber \\
a^2 \frac{\partial \sigma}{\partial x^i} & \sim 0 , \qquad 
a \displaystyle{\left(\frac{\partial \sigma}{\partial x^i} \right)^2} & 
\sim 0 \\ \nonumber
\end{eqnarray}

\noindent 
(with $ \sigma = \psi, \, \chi $) are neglected. The terms corresponding to 
the Kerr metric of the Ricci tensor components are also eliminated 
(see Appendix).

\noindent 
To solve the remaining terms of $ R_{i j} = 0 $, let propose the following 
Ansatz 

\begin{eqnarray}
\label{Anstaz}
\psi & = & \frac{q}{r^3} P_2 + \alpha \frac{M q}{r^4} P_2 \\ 
\chi & = & \frac{q P_2}{r^3} 
+ \frac{M q}{r^4} (\beta_1 + \beta_2 P_2 + \beta_3 P_2^2)
+ \frac{q^2}{r^6} (\beta_4 + \beta_5 P_2 + \beta_6 P_2^2 + \beta_7 P_2^3) , 
\nonumber
\end{eqnarray}

\noindent 
where $ q $ represents the quadrupole parameter, and 
$ P_2 = (3 \cos^2{\theta} - 1)/{2} $. Substituting this Ansatz into 
the Ricci tensor components, we get a set of linear equations for these 
constants $ \alpha $, and $ \beta_n $ ($ n = 1, \, \dots, \, 7 $). After 
solving these linear equations, the constants are

\begin{eqnarray}
\alpha & = & 3 \nonumber \\
\beta_1 & = & - \frac{1}{3} \nonumber \\
\beta_2 & = & \beta_3 = \frac{5}{3} \nonumber \\
\beta_4 & = & \frac{2}{9} \\
\beta_5 & = & - \frac{2}{3} \nonumber \\
\beta_6 & = & - \frac{7}{3} \nonumber \\
\beta_7 & = & \frac{25}{9} \nonumber
\end{eqnarray}

\newpage

\noindent 
From (\ref{superkerr}), the metric components reads

\begin{eqnarray}
\label{metriccomp}
g_{t t} & = & \frac{{\rm e}^{- 2 \psi}}{\rho^2} [a^2 \sin^2{\theta} - \Delta] 
\nonumber \\
g_{t \phi} & = & \frac{a}{\rho^2} [\Delta - (r^2 + a^2)] \sin^2{\theta} 
= - \frac{2 J r}{\rho^2} \sin^2{\theta} \\
g_{r r} & = & \rho^2 \frac{{\rm e}^{2 \chi}}{\Delta} \nonumber \\
g_{\theta \theta} & = & \rho^2 {\rm e}^{2 \chi} \nonumber \\
g_{\phi \phi} & = & \frac{{\rm e}^{2 \psi}}{\rho^2} 
[(r^2 + a^2)^2 - a^2 \Delta \sin^2{\theta}] \sin^2{\theta} .
\nonumber 
\end{eqnarray}

\noindent 
It was checked by means of a REDUCE program that the proposed metric is valid 
up to the order $ O(a q^2, \, a^2 q, \, M a q, \, M q^2, \, M^2 q, \, q^3) $.

\section{Comparison to the Hartle-Thorne Metric}
\label{sec:HT}

\noindent
In order to establish if our metric does really represent the gravitational 
field of an astrophysical object, we should show that it is possible to 
construct an interior solution, which can appropriately be matched with the
exterior solution. For this purpose, we employ the exterior HT spacetime 
\cite{HT,QM}. The HT metric describes the exterior of any slowly and rigidly 
rotating, stationary and axially symmetric body. It is an approximate solution 
of vacuum EFE. It has three parameters: mass $ M $, spin $ J $ and 
quadrupole-moment $ Q $. The accuracy of this spacetime is given with up to 
the second order terms in the body's angular momentum, and first order in its 
quadrupole moment. The HT solution is given by

\begin{equation} 
\label{ht} 
d {s}^2 = {g}_{tt} d t^2 + g_{rr} d r^2 +
{g}_{\theta \theta} d \theta^2 + {g}_{\theta \theta} \sin^2{\theta} d \phi^2
+ 2 {g}_{t \phi} d t d \phi ,
\end{equation}

\noindent
with  metric components

\begin{eqnarray}
\label{htcomp}
{g}_{tt} & = & - \left(1 - 2 U \right) [1 + 2 K_1 P_2(\cos{\theta})]
- 2 \frac{J^2}{r^4} (2 \cos^2{\theta} - 1) , \nonumber \\
g_{t \phi} & = & - 2 \frac{J}{r} \sin^2{\theta} , \\
g_{rr} & = & \frac{1}{1 - 2 U}
\left[1 - 2 K_2 P_2 (\cos{\theta}) - \frac{2}{1 - 2 U} \frac{J^2}{r^4} \right] ,
\nonumber \\
g_{\theta \theta} & = & r^2 [1 - 2 K_3 P_2 (\cos{\theta})] , \nonumber \\
g_{\phi \phi} & = & g_{\theta \theta} \sin^2{\theta} , \nonumber
\end{eqnarray}

\noindent
where

\begin{eqnarray}
\label{funcs}
K_1 & = & \frac{J^2}{M r^3} (1 + U)
+ \frac{5}{8} \left(\frac{q}{M^3} - \frac{J^2}{M^4} \right)
Q^{2}_{2} \left(\frac{r}{M} - 1 \right) ,
\nonumber \\
K_2 & = & K_1 - 6 \frac{J^2}{r^4} , \nonumber \\
K_3 & = & \left(K_1 + \frac{J^2}{r^4} \right)
+ \frac{5}{4} \left(\frac{q}{M^3} - \frac{J^2}{M^4} \right)
\frac{U}{\sqrt{1 - 2 U}} Q^{1}_{2} \left(\frac{r}{M} - 1 \right) , \nonumber
\end{eqnarray}

$$ U = \frac{M}{r} . $$

\noindent
The functions $ Q^{1,2}_{2} $  are
associated Legendre polynomials of the second kind

$$ Q^{1}_{2} = \sqrt{x^2 - 1} \left(\frac{3}{2} x
\ln{\left(\frac{1 + x}{1 - x} \right)}
- \frac{(3 x^2 - 2)}{(x^2 - 1)} \right) , $$

$$ Q^{2}_{2} = ({x^2 - 1}) \left(\frac{3}{2}
\ln{\left(\frac{1 + x}{1 - x} \right)}
- \frac{(3 x^3 - 5 x)}{(x^2 - 1)^2} \right) . $$

\noindent
A Taylor expantion of the metric components (\ref{htcomp}) 
til the second order of $ J, \, M $ and $ q $ leads to 

\begin{eqnarray}
\label{htexp}
g_{tt} & = & - \left(1 - 2 U + 2 \frac{q}{r^3} P_2 + 2 \frac{M q}{r^4} P_2 
+ 2 \frac{q^2}{r^6} P_2^2 - \frac{2}{3} \frac{J^2}{r^4} (2 P_2 + 1) \right) 
\nonumber \\
g_{t \phi} & = & - 2 \frac{J}{r} \sin^2{\theta} \\
g_{rr} & = & 1 + 2 U + 4 U^2 - 2 \frac{q}{r^3} P_2 - 10 \frac{M q}{r^4} P_2 
+ \frac{1}{12} \frac{q^2}{r^6} [8 P_2^2 - 16 P_2 + 77] 
+ 2 \frac{J^2}{r^4} (8 P_2 - 1) \nonumber \\
g_{\theta \theta} & = & r^2 \left(1 - 2 \frac{q}{r^3} P_2 - 5 \frac{M q}{r^4} P_2
+ \frac{1}{36} \frac{q^2}{r^6} [44 P_2^2 + 8 P_2 - 43]
+ \frac{J^2}{r^4} P_2 \right) \nonumber \\
g_{\phi \phi} & = & r^2 \sin^2{\theta} \left(1 - 2 \frac{q}{r^3} P_2 
- 5 \frac{M q}{r^4} P_2 + \frac{1}{36} \frac{q^2}{r^6} [44 P_2^2 + 8 P_2 - 43]
+ \frac{J^2}{r^4} P_2 \right) , \nonumber
\end{eqnarray}

\noindent
where we have added the second order terms of the quadrupole moment obtained by 
Frutos and Soffel \cite{Frutos3}.  

\noindent 
Now, let us expand in Taylor series the metric components (\ref{metriccomp}) 
til the second order of $ a, \, J, \, M $ and $ q $, the result is

\begin{eqnarray}
\label{postnewton}
g_{t t} & = & - \left(1 - 2 \frac{M}{r} + 2 \frac{M a^2}{r^3} \cos^2{\theta} 
- 2 \frac{q}{r^3} P_2 - 2 \frac{M q}{r^4} P_2 + 2 \frac{q^2}{r^6} P_2^2 \right) 
\nonumber \\
g_{t \phi} & = & - 2 \frac{J}{r} \sin^2{\theta} \\
g_{rr} & = & 1 + 2 \frac{M}{r} + 4 \frac{M^2}{r^2} 
- \frac{a^2}{r^2} \sin^2{\theta} 
- 2 \frac{M a^2}{r^3} (1 + \sin^2{\theta}) 
- 4 \frac{M^2 a^2}{r^4} (2 + \sin^2{\theta}) \nonumber \\
& + & 2 \frac{q}{r^3} P_2 + \frac{2}{3} \frac{M q}{r^4} (5 P_2^2 + 11 P_2 - 1) 
+ \frac{2}{9} \frac{q^2}{r^6} (25 P^3_2 - 12 P^2_2 - 6 P_2 + 2) \nonumber \\
g_{\theta \theta} & = & r^2 \left(1 + \frac{a^2}{r^2} \cos^2{\theta} 
+ 2 \frac{q}{r^3} P_2 + \frac{2}{3} \frac{M q}{r^4} (5 P_2^2 + 5 P_2 - 1) 
+ \frac{2}{9} \frac{q^2}{r^6} (25 P_2^3 - 12 P_2^2 - 6 P_2 + 2) \right)
\nonumber \\
g_{\phi \phi} & = & r^2 \sin^2{\theta} \left(1 + \frac{a^2}{r^2} 
+ 2 \frac{M a^2}{r^3} \sin^2{\theta}
+ 2 \frac{q}{r^3} P_2 + 6 \frac{M q}{r^4} P_2 + 2 \frac{q^2}{r^6} P_2^2 \right) .
\nonumber
\end{eqnarray}

\noindent
Comparing these results with the ones obtained by Frutos and Soffel 
\cite{Frutos3} for the ER metric, we note that both metric are the same if 
one neglects rotation and changes $ q \rightarrow 2 q / 15 $. 
Our metric corresponds to a rotating ER spacetime at this level of 
approximation.

\noindent
To compare our spacetime with the HT metric, we have to find a transformation 
that converts our metric (\ref{postnewton}) into the HT one (\ref{htexp}). 
The following transformation converts the Kerr-like truncated metric 
(\ref{postnewton}) into the improved HT spacetime (\ref{htexp}) changing 
$ q \rightarrow M a^2 - q $, at the same level of approximation.

\begin{eqnarray}
\label{trans}
r & = & R \left[1 + \frac{M q}{R^4} f_1 + \frac{q^2}{R^6} f_2 
+ \frac{a^2}{R^2} \left({h_1} + \frac{M}{R} h_2 + \frac{M^2}{R^2} h_3 \right) 
\right] \\
\theta & = & \Theta + \frac{M q}{R^4} g_1 + \frac{q^2}{R^6} g_2 
+ \frac{a^2}{R^2} \left({h_4} + \frac{M}{R} h_5 \right) , \nonumber
\end{eqnarray}

\noindent
where

\begin{eqnarray}
\label{functs}
f_1 & = & (\alpha_1 + \alpha_2 P_2 + \alpha_3 P_2^2) \nonumber \\
f_2 & = & (\alpha_4 + \alpha_5 P_2 + \alpha_6 P_2^2 + \alpha_7 P_2^3) 
\nonumber \\
g_1 & = & (\eta_1 + \eta_2 P_2 + \eta_3 P_2^2) \cos{\Theta} \sin{\Theta} 
\nonumber \\
g_2 & = & (\eta_4 + \eta_5 P_2 + \eta_6 P_2^2) \cos{\Theta} \sin{\Theta} \\
h_1 & = & \gamma_1 + \gamma_2 \cos^2{\Theta} \nonumber \\
h_2 & = & \gamma_3 + \gamma_4 \cos^2{\Theta} \nonumber \\
h_3 & = & \gamma_5 + \gamma_6 \cos^2{\Theta} \nonumber \\
h_4 & = & \gamma_7 \cos{\Theta} \sin{\Theta} \nonumber \\
h_5 & = & \gamma_8 \cos{\Theta} \sin{\Theta} . \nonumber 
\end{eqnarray}

\noindent
The constant are given by

\begin{center}
\begin{tabular}{ccc}
$ \alpha_1 = -1/9    $ & $ \eta_1 = 1/3   $ & $ \gamma_1 = -1/2  $ \\
$ \alpha_2 = - 4/9   $ & $ \eta_2 = - 5/6 $ & $ \gamma_2 = 1/2   $ \\
$ \alpha_3 = 5/9     $ & $ \eta_3 = 0     $ & $ \gamma_3 = - 1/2 $ \\
$ \alpha_4 = - 43/72 $ & $ \eta_4 = 0     $ & $ \gamma_4 = 1/2   $ \\
$ \alpha_5 = 0       $ & $ \eta_5 = 1/3   $ & $ \gamma_5 = 1     $ \\
$ \alpha_6 = - 1/3   $ & $ \eta_6 = - 5/6 $ & $ \gamma_6 = - 3   $ \\
$ \alpha_7  = 5/9    $ & $                $ & $ \gamma_7 = - 1/2 $ \\
                       &                    & $ \gamma_8 = - 1   $ .
\end{tabular}
\end{center}

\noindent
Since our expanded Kerr-like metric can be transformed to the improved HT 
spacetime, it is possible to construct an interior metric that could be matched 
to our exterior spacetime. It can be considered as an improvement 
of the HT spacetime.

\section{Comparison to other Stationary Metrics}
\label{sec:QMMN}

\noindent
There are many other stationary metrics. We concentrate on the Quevedo-Mashhoon 
\cite{QM,Quevedo} and the Manko-Novikov \cite{Manko} ones. At first glance, 
our metrics is not the same as these ones, because the rotational term $ W $ 
(\ref{potential}) has no quadrupole perturbation. To see if these metrics are 
the same, one has to compare the multipole structure. The first Ernst potential 
for metric (\ref{superkerr}) is \cite{Ernst}

\begin{eqnarray}
\label{Ernst1}
{\cal E} = f + i \Omega ,
\end{eqnarray}

\noindent
where $ f = V_K {\rm e}^{- 2 \psi} $ and $ \Omega $ is the twist scalar. 
To get this scalar, we have to solve the following equation 

\begin{eqnarray}
\label{dtwist}
\partial_{\alpha} \Omega & = & 
\varepsilon_{\alpha \beta \mu \nu} k^{\beta} \nabla^{\mu} k^{\nu} ,
\end{eqnarray}

\noindent
where $ k^{\beta} $ is the Killing vector, $ \nabla^{\mu} $ is the contravariant 
derivative and

\begin{eqnarray}
\label{eps}
\varepsilon_{\alpha \beta \mu \nu} = \sqrt{- g} \epsilon_{\alpha \beta \mu \nu} 
\end{eqnarray}

\noindent
Taking the Killing vector as in the Kerr metric 
$ k^{\beta} = (1, \,0, \,0, \,0) $. The result of (\ref{dtwist}) is given by 

\begin{eqnarray}
\label{twist}
\Omega & = & - \frac{2 J}{\rho^2} \cos{\theta} .
\end{eqnarray}

\noindent
This twist is the same as for the Kerr spacetime. Now, the second Ernst 
potential is \cite{Ernst}

\begin{eqnarray}
\label{Ernst2}
\xi & = & \frac{1 + {\cal E}}{1 - {\cal E}} = 
\frac{1 + f + i \Omega}{1 - (f + i \Omega)}
\end{eqnarray}

\noindent
One can show that this Ernst potential and its inverse are solutions of the 
Ernst equation. 

\begin{equation}
\label{Ernst}
({\xi} {\xi}^{\star} - 1) \nabla^2 {\xi} = 2 {\xi}^{\star} [\nabla {\xi}]^2 .
\end{equation}

\noindent
For the sake of calculating the relativistic multipole 
moments, it is better to employ the inverse potential \cite{Fodor}. 
Moreover, it is easier to calculate them using prolate spheroidal coordinates 
$ (t, \, x, \, y, \, \phi) $. The transformation to these coordinates is 
achieved by means of 

\begin{eqnarray}
\label{prolate}
\sigma x & = & r - M \\
y & = & \cos{\theta} , \nonumber 
\end{eqnarray}

\noindent
where $ \sigma^2 = M^2 - a^2 $.

\noindent
The potential $ V_K $, the twist scalar $ \Omega $ and the potential $ \psi $ 
are

\begin{eqnarray}
\label{ptransf}
V_K & = & \frac{(\sigma x)^2 - M^2 + (a y)^2}{(\sigma x + M)^2 + (a y)^2} 
\nonumber \\
\Omega & = & - \frac{2 J y}{(\sigma x + M)^2 + (a y)^2} \\
\psi & = & \frac{q P_2(y)}{(\sigma x + M)^3} 
\left(1 + \frac{3 M}{\sigma x + M} \right) , \nonumber
\end{eqnarray}

\noindent
where $ P_2(y) = (3 y^2 - 1) / 2 $.

\noindent
The procedure to get the relativistic multipole moments is the following 
\cite{Fodor}

\begin{enumerate}
\item employ the inverse Ernst potential $ \xi^{-1} $, 
\item set $ y = \cos{\theta} = 1 $ into $ \xi^{-1} $, 
\item change $ \sigma x \rightarrow 1 / z $ into $ \xi^{-1} $, 
\item expand in Taylor series of $ z $ the inverse Ernst potential, 
and finally, 
\item use the Fodor-Hoenselaers-Perj\'es (FHP) formulae \cite{Fodor}.
\end{enumerate}

\noindent
To obtain the multipole moment, we wrote a REDUCE program with the latter 
recipe. The first six mass and first five spin moments are 

\begin{eqnarray}
\label{sk}
{\cal M}_0 & = & M \nonumber \\
{\cal S}_1 & = & J = M a \nonumber \\
{\cal M}_2 & = & Q - M a^2 \nonumber \\
{\cal S}_3 & = & - M a^3 \nonumber \\
{\cal M}_4 & = & M a^4 \nonumber \\
{\cal S}_5 & = & M a^5 \nonumber \\
{\cal M}_6 & = & - M a^6 \nonumber \\
{\cal S}_7 & = & - M a^7 \nonumber \\
{\cal M}_8 & = & M a^8 \nonumber \\
{\cal S}_9 & = & M a^9 \nonumber \\
{\cal M}_{10} & = & - M a^{10} . \nonumber 
\end{eqnarray}

\noindent
A direct comparison of these multipole moments with the corresponding ones 
of QM \cite{Frutos4,Quevedo} and MN \cite{Manko} gives that 
the octupole $ {\cal S}_3 $ is different. Then, all these spacetimes are 
non-isometric. Moreover, it is clear that the only difference with the Kerr 
metric is that our metric has another term in quadrupole moment.

\section{Conclusion}

\noindent
Our metric was obtained solving the EFE perturbatively. The Lewis metric with 
the modified potentials from the Kerr spacetime was used. This metric has 
three parameters $ m $, $ a $ and $ q $ representing the mass, the rotation 
parameter and the quadrupole, respectively. It is valid until including 
$ m q $ and $ q^2 $ orders. This spacetime contains the Kerr metric, 
the expanded ER spacetime and an improvement of the HT metric, since as we have 
seen our expanded version correspond to a HT-like expanded spacetime until 
including $ m q $, $ q^2 $, and $ J^2 $ orders.

\noindent
The form of our expanded metric suggests that it is possible to construct an 
interior solution, because it can be transformed to the improved HT spacetime. 
It is known that the approximate exterior HT metric is coupled to the interior 
HT one. This gives meaning to our results. Our spacetime may represent 
the approximative spacetime of a rotating deformed object. Moreover, 
we improved the HT metric including the second order of the quadrupole moment 
accuracy. Furthermore, it seems that by means of our perturbation procedure, 
one could improve our metric to include more terms to a desirable accuracy.

\noindent
Moreover, the relativistic multipole moments were calculated to show that our 
spacetime was not isometric with the QM and the MN metrics. Our metric has a 
simple form and its multipole structure is Kerr-like, the only difference is 
that it has mass quadrupole. 

\noindent
This metric has potentially many applications because it could be employ 
as spacetime for real rotating astrophysical objects in a simple manner. 
Besides, it is easier to implement computer programs to apply this metric, 
because it maintains the simpleness of the Kerr metric. As an example of 
possible applications, the influence of the quadrupole moment in the light 
propagation and the light cone structure of this spacetime could be 
investigated using this Kerr-like spacetime. 



\section*{Appendix}

\noindent
The non-null Ricci tensor components for the metric (\ref{superkerr})  
(here $ V, \, W, \, X, \, Y, \, Z $ do not have the subscript $ K $ and 
$ R^{K}_{i j} $ refers to the Ricci tensor components of the Kerr metric) 
are given by

\begin{eqnarray}
R_{t t} & = & \frac{{\rm e}^{2 (\psi + 2 \chi)}}{4 \rho^2 X^2 Y^2} \left(
- 4 V^2 X^2 Y Z \frac{\partial^2 \psi}{\partial \theta^2} 
- 4 V W^2 X^2 Y \frac{\partial^2 \psi}{\partial \theta^2} \right. \nonumber \\
& + & \left. 
8 V W^2 X^2 Y \left(\frac{\partial \psi}{\partial \theta} \right)^2 
- 2 V X^2 Y Z \frac{\partial \psi}{\partial \theta} 
\frac{\partial V}{\partial \theta} 
- 8 W^2 X^2 Y \frac{\partial \psi}{\partial \theta} 
\frac{\partial V}{\partial \theta} \right. \nonumber \\
& + & \left. 4 V W X^2 Y \frac{\partial \psi}{\partial \theta} 
\frac{\partial W}{\partial \theta} 
- 2 V^2 X Y Z \frac{\partial \psi}{\partial \theta} 
\frac{\partial X}{\partial \theta} 
- 2 V W^2 X Y \frac{\partial \psi}{\partial \theta} 
\frac{\partial X}{\partial \theta} 
\right. \nonumber \\
& + & \left. 2 V^2 X^2 Z \frac{\partial \psi}{\partial \theta} 
\frac{\partial Y}{\partial \theta} 
+ 2 V W^2 X^2 \frac{\partial \psi}{\partial \theta} 
\frac{\partial Y}{\partial \theta} 
- 2 V^2 X^2 Y \frac{\partial \psi}{\partial \theta} 
\frac{\partial Z}{\partial \theta} \right. \nonumber \\
& - & \left. 4 V^2 X Y^2 Z \frac{\partial^2 \psi}{\partial r^2}  
- 4 V W^2 X Y^2 \frac{\partial^2 \psi}{\partial r^2} 
+ 8 V W^2 X Y^2 \left(\frac{\partial \psi}{\partial r} \right)^2 
\right. \nonumber \\
& - & \left. 2 V X Y^2 Z \frac{\partial \psi}{\partial r} 
\frac{\partial V}{\partial r} 
- 8 W^2 X Y^2 \frac{\partial \psi}{\partial r} \frac{\partial V}{\partial r} 
+ 4 V W X Y^2 \frac{\partial \psi}{\partial r} \frac{\partial W}{\partial r} 
\right. \nonumber \\
& + & \left. 2 V^2 Y^2 Z \frac{\partial \psi}{\partial r} 
\frac{\partial X}{\partial r} 
+ 2 V W^2 Y^2 \frac{\partial \psi}{\partial r} \frac{\partial X}{\partial r} 
- 2 V^2 X Y Z \frac{\partial \psi}{\partial r} \frac{\partial Y}{\partial r} 
\right. \nonumber \\
& - & \left. 2 V W^2 X Y \frac{\partial \psi}{\partial r} 
\frac{\partial Y}{\partial r} 
- 2 V^2 X Y^2 \frac{\partial \psi}{\partial r} \frac{\partial Z}{\partial r} 
\right) + R^{K}_{t t}{{\rm e}^{2 (\psi + 2 \chi)}} \nonumber 
\end{eqnarray}


\begin{eqnarray}
R_{t \phi} & = & \frac{{\rm e}^{- 2 \chi}}{4 \rho^2 X^2 Y^2} \left(
- 4 W X^2 Y Z \frac{\partial \psi}{\partial \theta} 
\frac{\partial V}{\partial \theta} 
+ 4 V W X^2 Y \frac{\partial \psi}{\partial \theta} 
\frac{\partial Z}{\partial \theta} \right. \nonumber \\ 
& + & \left. 
8 V W X^2 Y Z \left(\frac{\partial \psi}{\partial \theta} \right)^2 
+ 8 V W X Y^2 Z \left(\frac{\partial \psi}{\partial r} \right)^2 
\right. \nonumber \\ 
& - & \left. 4 W X Y^2 Z \frac{\partial \psi}{\partial r} 
\frac{\partial V}{\partial r} 
+ 4 V W X Y^2 \frac{\partial \psi}{\partial r} \frac{\partial Z}{\partial r} 
\right) + R^{K}_{t \phi}{{\rm e}^{- 2 \chi}} \nonumber 
\end{eqnarray}

\newpage

\begin{eqnarray}
R_{r r} & = & \frac{1}{4 \rho^4 X Y^2} \left(
4 V X Y^2 Z^2 \frac{\partial \psi}{\partial r} \frac{\partial V}{\partial r} 
+ 4 W^2 X Y^2 Z \frac{\partial \psi}{\partial r} \frac{\partial V}{\partial r}
\right. \nonumber \\  
& - & \left. 4 V^2 X Y^2 Z \frac{\partial \psi}{\partial r} 
\frac{\partial Z}{\partial r} 
- 8 V^2 X Y^2 Z^2 \left(\frac{\partial \psi}{\partial r} \right)^2 
- 8 V W^2 X Y^2 Z \left(\frac{\partial \psi}{\partial r} \right)^2 
\right. \nonumber \\ 
& - & \left. 4 V W^2 X Y^2 \frac{\partial \psi}{\partial r} 
\frac{\partial Z}{\partial r} 
- 4 V^2 X^2 Y Z^2 \frac{\partial^2 \chi}{\partial \theta^2} 
- 8 V W^2 X^2 Y Z \frac{\partial^2 \chi}{\partial \theta^2} \right. 
\nonumber \\ 
& - & \left. 4 W^4 X^2 Y \frac{\partial^2 \chi}{\partial \theta^2} 
- 2 V X^2 Y Z^2 \frac{\partial \chi}{\partial \theta} 
\frac{\partial V}{\partial \theta} 
- 2 W^2 X^2 Y Z \frac{\partial \chi}{\partial \theta} 
\frac{\partial V}{\partial \theta} \right. \nonumber \\ 
& - & \left. 4 V W X^2 Y Z \frac{\partial \chi}{\partial \theta} 
\frac{\partial W}{\partial \theta} 
- 4 W^3 X^2 Y \frac{\partial \chi}{\partial \theta} 
\frac{\partial W}{\partial \theta} 
- 2 V^2 X Y Z^2 \frac{\partial \chi}{\partial \theta} 
\frac{\partial X}{\partial \theta} \right. \nonumber \\ 
& - & \left. 4 V W^2 X Y Z \frac{\partial \chi}{\partial \theta} 
\frac{\partial X}{\partial \theta} 
- 2 W^4 X Y \frac{\partial \chi}{\partial \theta} 
\frac{\partial X}{\partial \theta} 
+ 2 V^2 X^2 Z^2 \frac{\partial \chi}{\partial \theta} 
\frac{\partial Y}{\partial \theta} \right. \nonumber \\ 
& + & \left. 4 V W^2 X^2 Z \frac{\partial \chi}{\partial \theta} 
\frac{\partial Y}{\partial \theta} 
+ 2 W^4 X^2 \frac{\partial \chi}{\partial \theta} 
\frac{\partial Y}{\partial \theta} 
- 2 V^2 X^2 Y Z \frac{\partial \chi}{\partial \theta} 
\frac{\partial Z}{\partial \theta} 
\right. \nonumber \\ 
& - & \left. 2 V W^2 X^2 Y \frac{\partial \chi}{\partial \theta} 
\frac{\partial Z}{\partial \theta} 
- 4 V^2 X Y^2 Z^2 \frac{\partial^2 \chi}{\partial r^2} 
- 8 V W^2 X Y^2 Z \frac{\partial^2 \chi}{\partial r^2} \right. \nonumber \\ 
& - & \left. 4 W^4 X Y^2 \frac{\partial^2 \chi}{\partial r^2} 
+ 2 V X Y^2 Z^2 \frac{\partial \chi}{\partial r} \frac{\partial V}{\partial r} 
+ 2 W^2 X Y^2 Z \frac{\partial \chi}{\partial r} \frac{\partial V}{\partial r} 
\right. \nonumber \\ 
& + & \left. 4 V W X Y^2 Z \frac{\partial \chi}{\partial r} 
\frac{\partial W}{\partial r} 
+ 4 W^3 X Y^2 \frac{\partial \chi}{\partial r} \frac{\partial W}{\partial r} 
+ 2 V^2 Y^2 Z^2 \frac{\partial \chi}{\partial r} \frac{\partial X}{\partial r} 
\right. \nonumber \\ 
& + & \left. 4 V W^2 Y^2 Z \frac{\partial \chi}{\partial r} 
\frac{\partial X}{\partial r} 
+ 2 W^4 Y^2 \frac{\partial \chi}{\partial r} \frac{\partial X}{\partial r} 
- 2 V^2 X Y Z^2 \frac{\partial \chi}{\partial r} \frac{\partial Y}{\partial r} 
\right. \nonumber \\ 
& - & \left. 4 V W^2 X Y Z \frac{\partial \chi}{\partial r} 
\frac{\partial Y}{\partial r} 
- 2 W^4 X Y \frac{\partial \chi}{\partial r} \frac{\partial Y}{\partial r} 
+ 2 V^2 X Y^2 Z \frac{\partial \chi}{\partial r} \frac{\partial Z}{\partial r} 
\right. \nonumber \\ 
& + & \left. 2 V W^2 X Y^2 \frac{\partial \chi}{\partial r} 
\frac{\partial Z}{\partial r} \right) + R^{K}_{r r} \nonumber
\end{eqnarray}


\begin{eqnarray}
R_{r \theta} & = & \frac{1}{4 \rho^4 X Y} \left(
- 8 V^2 X Y Z^2 \frac{\partial \psi}{\partial \theta} 
\frac{\partial \psi}{\partial r} 
- 8 V W^2 X Y Z \frac{\partial \psi}{\partial \theta} 
\frac{\partial \psi}{\partial r} 
\right. \nonumber \\ 
& + & \left. 2 V X Y Z^2 \frac{\partial \psi}{\partial \theta} 
\frac{\partial V}{\partial r} 
+ 2 W^2 X Y Z \frac{\partial \psi}{\partial \theta} 
\frac{\partial V}{\partial r} 
- 2 V^2 X Y Z \frac{\partial \psi}{\partial \theta} 
\frac{\partial Z}{\partial r} 
\right. \nonumber \\ 
& - & \left. 2 V W^2 X Y \frac{\partial \psi}{\partial \theta} 
\frac{\partial Z}{\partial r} 
+ 2 V X Y Z^2 \frac{\partial \psi}{\partial r} 
\frac{\partial V}{\partial \theta} 
+ 2 W^2 X Y Z \frac{\partial \psi}{\partial r} 
\frac{\partial V}{\partial \theta} 
\right. \nonumber \\ 
& - & \left. 2 V^2 X Y Z \frac{\partial \psi}{\partial r} 
\frac{\partial Z}{\partial \theta} 
- 2 V W^2 X Y \frac{\partial \psi}{\partial r} 
\frac{\partial Z}{\partial \theta} 
+ 2 V X Y Z^2 \frac{\partial \chi}{\partial \theta} 
\frac{\partial V}{\partial r} 
\right. \nonumber \\ 
& + & \left.  2 W^2 X Y Z \frac{\partial \chi}{\partial \theta} 
\frac{\partial V}{\partial r} 
+ 4 V W X Y Z \frac{\partial \chi}{\partial \theta} 
\frac{\partial W}{\partial r} 
+ 4 W^3 X Y \frac{\partial \chi}{\partial \theta} \frac{\partial W}{\partial r}
\right. \nonumber \\  
& + & \left. 2 V^2 X Y Z \frac{\partial \chi}{\partial \theta} 
\frac{\partial Z}{\partial r} 
+ 2 V W^2 X Y \frac{\partial \chi}{\partial \theta} 
\frac{\partial Z}{\partial r} 
+ 2 V X Y Z^2 \frac{\partial \chi}{\partial r} 
\frac{\partial V}{\partial \theta} 
\right. \nonumber \\  
& + & \left. 2 W^2 X Y Z \frac{\partial \chi}{\partial r} 
\frac{\partial V}{\partial \theta} 
+ 4 V W X Y Z \frac{\partial \chi}{\partial r} 
\frac{\partial W}{\partial \theta} 
+ 4 W^3 X Y \frac{\partial \chi}{\partial r} \frac{\partial W}{\partial \theta} 
\right. \nonumber \\ 
& + & \left. 2 V^2 X Y Z \frac{\partial \chi}{\partial r} 
\frac{\partial Z}{\partial \theta} 
+ 2 V W^2 X Y \frac{\partial \chi}{\partial r} 
\frac{\partial Z}{\partial \theta} \right) + R^{K}_{r \theta}\nonumber
\end{eqnarray}

\newpage

\begin{eqnarray}
R_{\theta \theta} & = & \frac{1}{4 \rho^4 X^2 Y} \left(
- 8 V^2 X^2 Y Z^2 \left(\frac{\partial \psi}{\partial \theta} \right)^2 
- 8 V W^2 X^2 Y Z \left(\frac{\partial \psi}{\partial \theta} \right)^2 
\right. \nonumber \\ 
& + & \left. 4 V X^2 Y Z^2 \frac{\partial \psi}{\partial \theta} 
\frac{\partial V}{\partial \theta} 
+ 4 W^2 X^2 Y Z \frac{\partial \psi}{\partial \theta} 
\frac{\partial V}{\partial \theta} 
- 4 V^2 X^2 Y Z \frac{\partial \psi}{\partial \theta} 
\frac{\partial Z}{\partial \theta} 
\right. \nonumber \\ 
& - & \left. 4 V W^2 X^2 Y \frac{\partial \psi}{\partial \theta} 
\frac{\partial Z}{\partial \theta} 
- 4 V^2 X^2 Y Z^2 \frac{\partial^2 \chi}{\partial \theta^2} 
- 8 V W^2 X^2 Y Z \frac{\partial^2 \chi}{\partial \theta^2} \right. 
\nonumber \\ 
& - & \left. 4 W^4 X^2 Y \frac{\partial^2 \chi}{\partial \theta^2} 
+ 2 V X^2 Y Z^2 \frac{\partial \chi}{\partial \theta} 
\frac{\partial V}{\partial \theta} 
+ 2 W^2 X^2 Y Z \frac{\partial \chi}{\partial \theta} 
\frac{\partial V}{\partial \theta} 
\right. \nonumber \\ 
& + & \left. 4 V W X^2 Y Z \frac{\partial \chi}{\partial \theta} 
\frac{\partial W}{\partial \theta} 
+ 4 W^3 X^2 Y \frac{\partial \chi}{\partial \theta} 
\frac{\partial W}{\partial \theta} 
- 2 V^2 X Y Z^2 \frac{\partial \chi}{\partial \theta} 
\frac{\partial X}{\partial \theta} 
\right. \nonumber \\ 
& - & \left. 4 V W^2 X Y Z \frac{\partial \chi}{\partial \theta} 
\frac{\partial X}{\partial \theta} 
- 2 W^4 X Y \frac{\partial \chi}{\partial \theta} 
\frac{\partial X}{\partial \theta} 
+ 2 V^2 X^2 Z^2 \frac{\partial \chi}{\partial \theta} 
\frac{\partial Y}{\partial \theta} 
\right. \nonumber \\ 
& + & \left. 4 V W^2 X^2 Z \frac{\partial \chi}{\partial \theta} 
\frac{\partial Y}{\partial \theta} 
+ 2 W^4 X^2 \frac{\partial \chi}{\partial \theta} 
\frac{\partial Y}{\partial \theta} 
+ 2 V^2 X^2 Y Z \frac{\partial \chi}{\partial \theta} 
\frac{\partial Z}{\partial \theta} 
\right. \nonumber \\ 
& + & \left. 2 V W^2 X^2 Y \frac{\partial \chi}{\partial \theta} 
\frac{\partial Z}{\partial \theta} 
- 4 V^2 X Y^2 Z^2 \frac{\partial^2 \chi}{\partial r^2} 
- 8 V W^2 X Y^2 Z \frac{\partial^2 \chi}{\partial r^2} \right. \nonumber \\ 
& - & \left. 4 W^4 X Y^2 \frac{\partial^2 \chi}{\partial r^2} 
- 2 V X Y^2 Z^2 \frac{\partial \chi}{\partial r} \frac{\partial V}{\partial r} 
- 2 W^2 X Y^2 Z \frac{\partial \chi}{\partial r} \frac{\partial V}{\partial r} 
\right. \nonumber \\ 
& - & \left. 4 V W X Y^2 Z \frac{\partial \chi}{\partial r} 
\frac{\partial W}{\partial r} 
- 4 W^3 X Y^2 \frac{\partial \chi}{\partial r} \frac{\partial W}{\partial r} 
+ 2 V^2 Y^2 Z^2 \frac{\partial \chi}{\partial r} \frac{\partial X}{\partial r} 
\right. \nonumber \\ 
& + & \left. 4 V W^2 Y^2 Z \frac{\partial \chi}{\partial r} 
\frac{\partial X}{\partial r} 
+ 2 W^4 Y^2 \frac{\partial \chi}{\partial r} \frac{\partial X}{\partial r} 
- 2 V^2 X Y Z^2 \frac{\partial \chi}{\partial r} \frac{\partial Y}{\partial r} 
\right. \nonumber \\ 
& - & \left. 4 V W^2 X Y Z \frac{\partial \chi}{\partial r} 
\frac{\partial Y}{\partial r} 
- 2 W^4 X Y \frac{\partial \chi}{\partial r} \frac{\partial Y}{\partial r} 
- 2 V^2 X Y^2 Z \frac{\partial \chi}{\partial r} \frac{\partial Z}{\partial r} 
\right. \nonumber \\ 
& - & \left. 2 V W^2 X Y^2 \frac{\partial \chi}{\partial r} 
\frac{\partial Z}{\partial r} \right) + R^{K}_{\theta \theta} \nonumber 
\end{eqnarray}


\begin{eqnarray}
R_{\phi \phi} & = & \frac{{\rm e}^{2 (\psi - \chi)}}{4 \rho^2 X^2 Y^2} \left(
- 4 V X^2 Y Z^2 \frac{\partial^2 \psi}{\partial \theta^2} 
- 4 W^2 X^2 Y Z \frac{\partial^2 \psi}{\partial \theta^2} 
\right. \nonumber \\ 
& - & \left. 
8 W^2 X^2 Y Z \left(\frac{\partial \psi}{\partial \theta} \right)^2 
- 2 X^2 Y Z^2 \frac{\partial \psi}{\partial \theta} 
\frac{\partial V}{\partial \theta} 
+ 4 W X^2 Y Z \frac{\partial \psi}{\partial \theta} 
\frac{\partial W}{\partial \theta} 
\right. \nonumber \\ 
& - & \left. 2 V X Y Z^2 \frac{\partial \psi}{\partial \theta} 
\frac{\partial X}{\partial \theta} 
- 2 W^2 X Y Z \frac{\partial \psi}{\partial \theta} 
\frac{\partial X}{\partial \theta} 
+ 2 V X^2 Z^2 \frac{\partial \psi}{\partial \theta} 
\frac{\partial Y}{\partial \theta} 
\right. \nonumber \\ 
& + & \left. 2 W^2 X^2 Z \frac{\partial \psi}{\partial \theta} 
\frac{\partial Y}{\partial \theta} 
- 2 V X^2 Y Z \frac{\partial \psi}{\partial \theta} 
\frac{\partial Z}{\partial \theta} 
- 8 W^2 X^2 Y \frac{\partial \psi}{\partial \theta} 
\frac{\partial Z}{\partial \theta} \right. \nonumber \\  
& - & \left. 4 V X Y^2 Z^2 \frac{\partial^2 \psi}{\partial r^2} 
- 4 W^2 X Y^2 Z \frac{\partial^2 \psi}{\partial r^2} 
- 8 W^2 X Y^2 Z \left(\frac{\partial \psi}{\partial r} \right)^2 
\right. \nonumber \\ 
& - & \left. 2 X Y^2 Z^2 \frac{\partial \psi}{\partial r} 
\frac{\partial V}{\partial r} 
+ 4 W X Y^2 Z \frac{\partial \psi}{\partial r} \frac{\partial W}{\partial r} 
+ 2 V Y^2 Z^2 \frac{\partial \psi}{\partial r} \frac{\partial X}{\partial r} 
\right. \nonumber \\ 
& + & \left. 2 W^2 Y^2 Z \frac{\partial \psi}{\partial r} 
\frac{\partial X}{\partial r} 
- 2 V X Y Z^2 \frac{\partial \psi}{\partial r} \frac{\partial Y}{\partial r} 
- 2 W^2 X Y Z \frac{\partial \psi}{\partial r} \frac{\partial Y}{\partial r} 
\right. \nonumber \\ 
& - & \left. 2 V X Y^2 Z \frac{\partial \psi}{\partial r} 
\frac{\partial Z}{\partial r} 
- 8 W^2 X Y^2 \frac{\partial \psi}{\partial r} \frac{\partial Z}{\partial r} 
\right) + R^{K}_{\phi \phi}{{\rm e}^{2 (\psi - \chi)}} \nonumber
\end{eqnarray}


\begin{thebibliography}{99}

\bibitem{Boshkayev}
{K.~Boshkayev, H.~Quevedo and R.~Ruffini.} 
{Gravitational field of compact objects in general relativity.} 
{Physical Review D}, {86}, 064043, 2012.

\bibitem{Carmeli}
{M.~Carmeli.}
{\em Classical Fields: General Relativity and Gauge Theory.}
World Scientific Publishing, Singapore, 2001.
{http://www.worldscientific.com/worldscibooks/10.1142/4843}

\bibitem{Ernst}
{F.~J.~Ernst.}
{New Formulation of the Axially Symmetric Gravitational Field Problem.}
{Physical Review}, {167}(5), 1175--1177, 1968.
{http://dx.doi.org/10.1103/PhysRev.167.1175}

\bibitem{Fodor}
{G.~Fodor, C.~Hoenselaers, and Z.~Perj\'es.} 
{Multipole moments of axisymmetric systems in relativity.} 
{Journal of Mathematical Physics}, {30}, 2252--2257, 1989. 
{http://dx.doi.org/10.1063/1.528551}

\bibitem{Frutos1}
{F.~Frutos-Alfaro, E.~Retana-Montenegro, I.~Cordero-Garc{\'{\i}}a, 
and J.~Bonatti-Gonz\'alez.} 
{Metric of a Slow Rotating Body with Quadrupole 
Moment from the Erez-Rosen Metric.}
{International Journal of Astronomy and Astrophysics}, {3}, 431--437, 2013. \\
{http://dx.doi.org/10.4236/ijaa.2013.34051}

\bibitem{Frutos2}
{F.~Frutos-Alfaro, P.~Montero-Camacho, M.~Araya, and J.~Bonatti-Gonz\'alez.}
{Approximate Metric for a Rotating Deformed Mass.}
{International Journal of Astronomy and Astrophysics}, {5}, 1--10, 2015.
{http://dx.doi.org/10.4236/ijaa.2015.51001}

\bibitem{FrutGrave}
{Francisco Frutos-Alfaro, Frank Grave, Thomas M\"ueller, Daria Adis}
{Wavefronts and Light Cones for Kerr Spacetimes.}
{Journal of Modern Physics}, 2012, {3}, 1882-1890, 2012. \\
{http://dx.doi.org/10.4236/jmp.2012.312237}
 
\bibitem{Frutos3}
{F.~Frutos-Alfaro and M.~Soffel.}
{On the Post-linear Quadrupole-Quadrupole Metric.}, 2015. \\
{http://arxiv.org/abs/1507.04264}

\bibitem{Frutos4}
{F.~Frutos-Alfaro and M.~Soffel.}
{Multipole moments of the generalized Quevedo-Mashhoon metric.}, 2016.
{http://arxiv.org/abs/1606.07173}

\bibitem{HT}
{J.~B.~Hartle and K.~S.~Thorne.}
{Slowly rotating relativistic stars. II.
Models for neutron stars and supermassive stars.}
{Astrophysical Journal}, {153}, 807--834, 1968. \\
{http://dx.doi.org/10.1086/149707}

\bibitem{Hearn}
{A.~C.~Hearn.}
{\em REDUCE} (User's and Contributed Packages Manual).
Konrad-Zuse-Zentrum f\"ur Informationstechnik, Berlin, 1999.
{http://www.reduce-algebra.com/docs/reduce.pdf}

\bibitem{Hernandez}
{W.~Hern\'andez.} 
{Material Sources for the Kerr Metric.} 
{Physical Review}, {159}, 1070--1072, 1967. 
{http://dx.doi.org/10.1103/PhysRev.159.1070}

\bibitem{HKX}
{C.~Hoenselaers, W.~Kinnersley and B.~C.~Xanthopoulos.} 
{Symmetries of the stationary Einstein-Maxwell equations. VI. 
Transformations which generate asymptotically flat spacetimes with arbitrary 
multipole moments.}
{Journal of Mathematical Physics}, {20}(12), 2530--2536, 1979. \\
{http://dx.doi.org/10.1063/1.523580}

\bibitem{Manko}
{V.~S.~Manko and I.~D.~Novikov.} 
{Generalizations of the Kerr and Kerr-Newman Metrics Possessing an Arbitrary 
Set of Mass-Multipole Moments.} 
{Classical and Quantum Gravity}, {9}, 2477-2487, 1992.
{http://dx.doi.org/10.1088/0264-9381/9/11/013}

\bibitem{Montero}
{P.~Montero-Camacho, F.~Frutos-Alfaro, C.~Guti\'errez-Chaves.}
{Slowly rotating Curzon-Chazy Metric.} 
{Revista de Matem\'atica} (Teor{\'{\i}}a y Aplicaciones) {22}(2), 
265--274, 2015. \\
{http://dx.doi.org/10.15517/rmta.v22i2.20833}  
({http://arxiv.org/abs/1405.2899})

\bibitem{QM}
{H.~Quevedo and B.~Mashhoon.}
{Generalization of Kerr spacetime.}
{Physical Review}, {43}(12), 3902--3906, 1991.
{http://dx.doi.org/10.1103/PhysRevD.43.3902}

\bibitem{Quevedo}
{H.~Quevedo.}
{Exterior and interior metrics with quadrupole moment.}
{General Relativity and Gravitation}, {43}(4), 1141--1152, 2011.
{http://dx.doi.org/10.1007/s10714-010-0940-5}

\bibitem{Thorne}
{K.~S.~Thorne.}
{Multipole expansions of gravitational radiation.}
{Reviews on Modern Physics}, {52}(2), 299--340, 1980.
{http://dx.doi.org/10.1103/RevModPhys.52.299}

\end{thebibliography}
\end{document}